\documentclass[jgrga]{agu2001}
\usepackage{amsmath}
\usepackage{amssymb}
\usepackage{graphicx}
\usepackage{bm}

\journalid{}
\articleid{}{}
\paperid{}
\cpright{}{}
\received{} \revised{} \accepted{} \published{}

\authorrunninghead{SWISDAK ET AL.}
\titlerunninghead{DIAMAGNETIC SUPPRESSION OF RECONNECTION}
\authoraddr{M. Swisdak,
IREAP, University of Maryland,
College Park, MD 20742-3511, USA,
(swisdak@glue.umd.edu)}

\begin{document}

\title{Diamagnetic Suppression of Component Magnetic Reconnection at
the Magnetopause}

\authors{M. Swisdak, \altaffilmark{1}
B. N. Rogers, \altaffilmark{2}
J. F. Drake, \altaffilmark{1}
and M. A. Shay\altaffilmark{1}}

\altaffiltext{1}
{IREAP, University of Maryland, College Park, MD, USA.}

\altaffiltext{2}{Department of Physics, Dartmouth College,
Hanover, NH, USA.}

\begin{abstract}
We present particle-in-cell simulations of collisionless magnetic
reconnection in a system (like the magnetopause) with a large density
asymmetry across the current layer.  In the presence of an ambient
component of the magnetic field perpendicular to the reconnection
plane the gradient creates a diamagnetic drift that advects the X-line
with the electron diamagnetic velocity.  When the relative drift
between the ions and electrons is of the order the Alfv\'en speed the
large scale outflows from the X-line necessary for fast reconnection
cannot develop and the reconnection is suppressed.  We discuss how
these effects vary with both the plasma $\beta$ and the shear angle of
the reconnecting field and discuss observational evidence for
diamagnetic stabilization at the magnetopause.
\end{abstract}

\begin{article}

\section{\label{intro}Introduction}
Magnetic reconnection, the breaking and reforming of magnetic field
lines with a concomitant transfer of energy from the field to the
surrounding plasma, is thought to drive such phenomena as solar
flares, magnetospheric substorms, and tokamak sawtooth crashes.
Despite the suspected linkage, the resistive magnetohydrodynamic (MHD)
model of reconnection due to {\it Sweet} [1958] and {\it Parker}
[1957] (the favored theoretical model for many years),
\nocite{parker57b,sweet58a} predicts reconnection rates in each of
these systems that are typically orders of magnitude too low to be
physically relevant.  The discrepancy arises because the Sweet-Parker
reconnection rate varies with the Spitzer resistivity which in turn
depends on particle interactions that are rare in these collisionless
plasmas.

Recent work suggests that terms important at small length scales but
usually ordered away in resistive MHD, notably the Hall term in Ohm's
law, lead to reconnection rates that are consistent with the
observations [{\it Shay et al.}, 1999]\nocite{shay99a}.  With the
addition of this new physics the speed of the electron flows away from
the reconnection site (the X-line) is no longer bounded by the ion
Alfv\'en speed, instead scaling inversely with the width of the
current layer formed by the reconnecting magnetic fields.  As a
result, the outward electron flux remains large even as the width,
controlled by non-ideal effects, becomes small.  Downstream from the
reconnection site the super-Alfv\'enic electrons slow to rejoin the
ions and expand outwards in a wide layer reminiscent of the {\it
Petschek} [1964] model\nocite{petschek64a}. This behavior has been
confirmed by a variety of numerical simulations: two-fluid, hybrid
(fluid electrons and particle ions), and full particle.  In the GEM
(Geospace Environment Modeling) reconnection challenge each of several
independent codes modeling an identical (two-dimensional) system found
features similar to those described above [{\it Birn et al.}, 2001 and
references therein]\nocite{birn01a}.  Laboratory experiments in the
relevant regimes are challenging, but recent results [{\it Brown},
1999; {\it Ji et al.}, 1999] \nocite{brown99a,ji99a} support some
aspects of this picture.

Although the GEM challenge established a possible mechanism for fast
collisionless reconnection, it did so for a relatively simple
geometry.  Experimental evidence suggests that in nature reconnection
is likely to be more complex.  For instance, at the Earth's dayside
magnetopause the magnetosphere, a region of low density but strong
magnetic field, abuts the magnetosheath and its high density but
weaker magnetic field.  Still, early single-point spacecraft
measurements of the magnetopause [{\it Paschmann et al.}, 1979; {\it
Sonnerup et al.}, 1981] \nocite{paschmann79a,sonnerup81a} detected
field signatures and particle distribution functions suggesting the
occurrence of reconnection.  A more recent study using data from
multiple spacecraft [{\it Phan et al.}, 2000] \nocite{phan00a} found,
at least in one case, the bidirectional jets that are a reconnection
hallmark.  Similar cross-field pressure gradients occur in laboratory
fusion experiments where their associated diamagnetic drifts are
thought to control the onset of reconnection during the sawtooth crash
[{\it Levinton et al.}, 1994].\nocite{levinton94a}

The first theoretical explorations of asymmetric reconnection [{\it
Levy et al.}, 1964; {\it Petschek and Thorne}, 1967]
\nocite{levy64a,petschek67a} were based on incompressible MHD and
predict an outflow region comprising a combination of slow MHD shocks
and slow and intermediate MHD waves.  Two-dimensional computer
simulations using compressible MHD [{\it Hoshino and Nishida}, 1983;
{\it Scholer}, 1989], \nocite{hoshino83a,scholer89a} hybrid [{\it Lin
and Xie}, 1997; {\it Omidi et al.}, 1998; {\it Krauss-Varban et al.},
1999; {\it Nakamura and Scholer}, 2000] \nocite{lin97a, omidi98a,
kraussvarban99a, nakamura00a} and particle [{\it Okuda},
1993]\nocite{okuda93a} codes have addressed these predictions.  Common
features include the preferential growth of the magnetic island
(O-line) towards the magnetosheath and a pronounced density drop on
the magnetospheric side of the boundary layer.  Also, consistent with
the picture of {\it Levy et al.}  [1964]\nocite{levy64a}, the current
layer abutting the magnetosheath is generally stronger than that
bordering the magnetosphere and takes the form of a series of
discontinuities. There is disagreement, however, concerning the
precise form of these discontinuities, possibly because the
nonlocality associated with the motion of collisionless particles
parallel to the shock front makes the two-dimensional (2-D)
simulations more complicated than the 1-D models.  Simulations have
also been performed with an initial out-of-plane component of the
magnetic field (nonzero $B_y$ in GSM coordinates) [{\it Hoshino and
Nishida}, 1983; {\it Krauss-Varban et al.}, 1999; {\it Nakamura and
Scholer}, 2000]. This guide field tends to slow, but not completely
suppress, reconnection and alter the structure of the shock
transitions.

In our investigations we begin with a simple magnetopause equilibrium
model with an ambient pressure difference and perform two-dimensional
fully kinetic numerical simulations.  We reproduce the qualitative
features seen in previous work but also find that diamagnetic drifts
produced by the pressure gradient advect the X-line. When the
diamagnetic velocity, $\mathbf{v}_*=-(c/qnB^2)\bm{\nabla} p
\bm{\times} \mathbf{B}$ ($c$ is the speed of light; $q$, $n$, and $p$
are the species charge, density, and pressure; $\mathbf{B}$ is the
magnetic field), is comparable to the Alfv\'en speed, reconnection is
completely suppressed.  These diamagnetic effects have not been
previously reported in the context of magnetopause reconnection. They
are ordered out of the MHD model [{\it Hoshino and Nishida}, 1983] and
were not seen in the hybrid simulations, perhaps because the spatially
localized resistivity used to trigger fast reconnection effectively
prohibits X-line advection.  In spite of this, diamagnetic effects can
be significant at the magnetopause.  Taking $n \sim 10 \text{
cm}^{-3}$, $B \sim 10^{-4} \text{ G}$, $k_BT \sim 100 \text{ eV}$, and
gradients of order an inverse ion inertial length, one finds $v_*/v_A
\sim 1$.  In the fusion context, previous analytic work has suggested
that diamagnetic drifts can stabilize the tearing mode and lower the
rate of reconnection \nocite{biskamp81a, rogers95a}[{\it Biskamp},
1981; {\it Rogers and Zakharov}, 1995].  But since this work
considered a reduced MHD model (equivalent to the limit of a large
guide field) its applicability to magnetospheric applications has not
been established.

Unlike the configuration in the magnetotail, the magnetic field on
opposite sides of the magnetopause is usually not equal in magnitude
and anti-parallel.  Reconnection in such a system does not occur at a
unique spatial location, even in a model with a 1-D equilibrium, since
components of the magnetic field reverse direction at any location
across the current layer. Because of the possibility that reconnection
can occur at multiple spatial locations, it has been suggested that
the magnetopause magnetic field is stochastic [{\it Galeev et al.},
1986; {\it Lee et al.}, 1993]\nocite{galeev86a,lee93a}. The
exploration of reconnection in a 2-D model must then be carried out
with care since stability at a particular surface does not necessarily
imply stability at all surfaces. In exploring the impact of
diamagnetic drifts on reconnection we therefore must address both
stability at a single plane and explore which plane is expected to
dominate the dynamics of a full 3-D system. We suggest on the basis of
analytic arguments and simulations that the strongest reconnection
occurs at the surface where the reversed field components have equal
magnitudes.

In section \ref{compute} of the paper we present our computational scheme
and initial conditions.  Section \ref{simoverview} is an overview of
reconnection at the magnetopause, both without and with a diamagnetic
drift.  Section \ref{betasec} discusses how varying the strength of
the out-of-plane field changes the diamagnetic effects while
section \ref{plane} addresses the question of determining the dominant
reconnection plane.  Finally, in section \ref{conclusions} we summarize our
results and discuss their implications for understanding magnetic
reconnection at the magnetopause.

\section{\label{compute}Computational Methods}

Although diamagnetic drifts are present in two-fluid and hybrid
models, a complete calculation in the collisionless limit requires a
careful, and assumption-filled, treatment of the full pressure tensor.
To avoid this difficulty, we sacrifice the benefits of a fluid
simulation for a more computationally demanding, but mathematically
straightforward, full particle description.

\subsection{The Code}

The simulations are done with p3d, a massively parallel kinetic code
that can evolve up to $\sim\negmedspace 10^9$ particles on current
Cray T3E and IBM SP systems [{\it Zeiler et al.},
2002]\nocite{zeiler02a}.  The Lorentz equation of motion for each
particle is evolved by a Boris algorithm ($\mathbf{E}$ accelerates for
half a timestep, followed by a rotation of $\mathbf{v}$ by
$\mathbf{B}$, and then the other half-step acceleration by
$\mathbf{E}$).  The electromagnetic fields are advanced in time with
an explicit trapezoidal-leapfrog method using second-order spatial
derivatives.  Poisson's equation constrains the system; if $\bm{\nabla
\cdot} \mathbf{E} \neq 4\pi \rho $, a multigrid algorithm recursively
corrects the electric field.  Although the code permits other choices,
we work with fully periodic boundary conditions.

The code is written in normalized units: lengths to the ion inertial
length $c/\omega_{pi} = d_i$, times to the inverse ion cyclotron
frequency $\Omega_{ci}^{-1}$, velocities to the Alfv\'en speed $v_A$,
masses to the ion mass $m_i$, and temperatures to $m_i v_A^2$.  Unless
otherwise noted, the asymptotic reconnecting field is assumed to have a
magnitude of $1$ on both sides of the magnetopause (the reason for
this choice will be made clear in section \ref{plane}) and the asymptotic
density in the magnetosheath is taken to be $1$.

To conserve computational resources, yet assure a sufficient
separation of spatial and temporal scales, we take the electron mass
to be $0.005 = 1/200$ and the speed of light to be $20$.  Our
simulations are performed in a box of $1024\times 1024$ gridpoints
that is $25.6$ on a side, corresponding to a grid scale of $0.025$ and
so $\approx\negmedspace 3$ gridpoints per electron inertial length.  A
typical timestep is $1.5\times 10^{-3}$.

\subsection{\label{inits}Initial Conditions}

Reconnection occurs in the $x-y$ plane in our coordinate system. For
reference, this is equivalent to the $z-x$ plane in GSM coordinates.
The initial equilibrium is constructed from two (modified) Harris
current sheets centered at $L_y/4$ and $3L_y/4$, where $L_y$ is the
box size in the $y$ direction; to allow periodicity the current is
parallel to $\mathbf{\hat{z}}$ in one sheet and anti-parallel in the
other. The current sheets produce a magnetic field
$B_x=\tanh[(y-L_y/4)/w_0]-\tanh[(y-3L_y/4)/w_0]-1$, where $w_0 =
0.5$. The density profile is similar, $n =
n_0(\tanh[(y-L_y/4)/w_0]-\tanh[(y-3L_y/4)/w_0])+n_{\text{min}}$,
where, unless otherwise specified, $n_0 = 0.45$ and $n_{\text{min}} =
0.1$.  This profile gives an asymptotic density of $n=1$ in the
magnetosheath and $0.1$ in the magnetosphere. The initial ion and
electron temperatures are $T_i = 2$ and $T_e = 1$, implying thermal
speeds of $\approx 1.4 \text{ and } 14$, respectively.  The guide
(out-of-plane) field $B_z$ is assigned a specific asymptotic value
$B_{z0}$ on the magnetosheath side and is calculated elsewhere by
assuming pressure balance.  Note that at the reversal surface, where
$B_x = 0$, pressure balance implies that $\partial_y n(T_i+T_e)
\propto B_z\partial_y B_z$, and hence a density gradient must be
accompanied by a gradient in $B_z$.  At $t=0$ a 5\% perturbation in
the magnetic flux function places the X- and O-line in each sheet.

Unlike a conventional Harris sheet, this initial state is not a
kinetic (Vlasov) equilibrium. Although it has been shown [{\it Quest
and Coroniti}, 1981]\nocite{quest81a} that the presence of a guide
field reduces the growth rate of the tearing mode instability in the
linear regime, our initial perturbation is large enough that the
system begins to reconnect nonlinearly before any other perturbations
become important.  We must also emphasize that our system
characterizes the magnetopause only in a rough approximation.  In
particular, magnetospheric plasma is typically not isothermal, but
instead has temperature gradients comparable in magnitude, but
anti-parallel, to those of the plasma density.  If this effect is
included the overall pressure gradient, as well as the diamagnetic
effects we discuss here, will be smaller.

\section{\label{simoverview}Simulation Overview}

In the simplest description of fast collisionless reconnection three
length scales are relevant: the system size and the ion and electron
inertial lengths. (Depending on the strength of the initial
out-of-plane field, the ion Larmor radius can also enter as a
parameter [{\it Kleva and Drake}, 1995]\nocite{kleva95a}.) The dynamic
couplings between these scales, which must exist for fast reconnection
to occur, are discussed in detail by \nocite{biskamp97a,shay98a}{\it
Biskamp et al.} [1997] and {\it Shay et al.} [1998]. At large scales
both species are frozen into the magnetic field and flow towards the
X-line with a velocity $v_{in}$ that is some fraction of the upstream
Alfv\'en speed.  Roughly $c/\omega_{pi}$ away from the current sheet
the inertia term in the ion equation of motion becomes important and
the ions, but not the electrons, decouple from the field.  Deflected
outwards, the ions flow Alfv\'enically away from the X-line while the
electrons travel further inward until they too decouple and travel
outward at velocities near the electron Alfv\'en speed.  Downstream,
the electrons slow to join the ionic outflow and reform an MHD
fluid. In what follows we show that diamagnetic drifts can disrupt
this structure by strongly altering the MHD outflows, thereby
inhibiting the coupling of the ions to the X-line and halting the
formation of a large-scale reconnection geometry.

\subsection{Anti-parallel Reconnection}\label{nodrift}

\begin{figure}
\noindent\includegraphics[width=20pc]{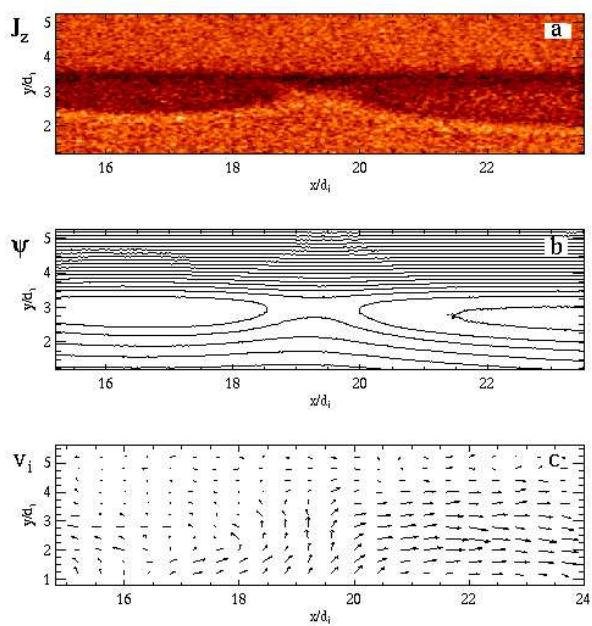}
\caption{\label{psifig} Reconnection in a magnetopause-like
configuration with no guide field. In each panel $t = 13.5
\Omega_{ci}^{-1}$.  (a) The out-of-plane current density $J_z$, (b)
the in-plane field lines, and (c) the ion velocity vectors, averaged
over a scale of $\sim d_i$. Note the flows associated with the X-line.
The longest vector corresponds to a velocity of $\sim 0.15 v_A$.}
\end{figure}

To document certain features and establish contact with previous work,
Figure \ref{psifig} shows the X-line structure from a magnetopause
simulation similar to those described by {\it Krauss-Varban et al.}
[1999] \nocite{kraussvarban99a} and {\it Nakamura and Scholer}
[2000]\nocite{nakamura00a}.  The upper section of each panel ($y/d_i >
3.2$) contains low density plasma and a large magnetic field like the
magnetosphere, while the lower section has a high density plasma and a
weak field like the magnetosheath. There are a few differences between
the initial conditions for this simulation and the case with a guide
field discussed in section \ref{inits} (and used in the rest of this paper).
The reversed field is given by a shifted hyperbolic tangent such that
the magnitudes of the asymptotic fields are unequal, $B_x^{sp} = 1.5$
and $B_x^{sh}=0.5$, where magnetosphere and magnetosheath quantities
are denoted with the superscripts ``sp'' and ``sh'', respectively.
The guide field $B_z$ is zero.  Pressure balance then fixes the
density profile as a gradient from $1$ to $2/3$ across the current
layer with a relative maxima where $B_x = 0$.

The general features of Figure \ref{psifig} are consistent with
earlier calculations. One consequence of the asymmetry in field
strengths across the magnetopause is the evident tendency for the
magnetic island to grow preferentially in the direction of the
magnetosheath.  Since equal amounts of flux must reconnect from either
side in a given time, $B^{sp} v_{in}^{sp} = B^{sh} v_{in}^{sh}$, and
so $v_{in}^{sh} = (B^{sp}/B^{sh})v_{in}^{sp}\gg v_{in}^{sp}$. The
larger inflow on the magnetosheath side of the reversal surface is
evident in the ion velocity vectors shown in Figure \ref{psifig}c.
With the stronger inflow, the (frozen-in) magnetosheath field is more
strongly advected toward the X-line than its magnetospheric
counterpart, leading to the wider opening angle ({\it i.e.}, the
bulge) on the magnetosheath side of the current layer.  In earlier MHD
simulations it was shown that the asymmetry of the island did not
appear when the amplitudes of the reversed field on either side of the
magnetopause are equal [{\it Scholer}, 1989].

\begin{figure}
\noindent\includegraphics[width=20pc]{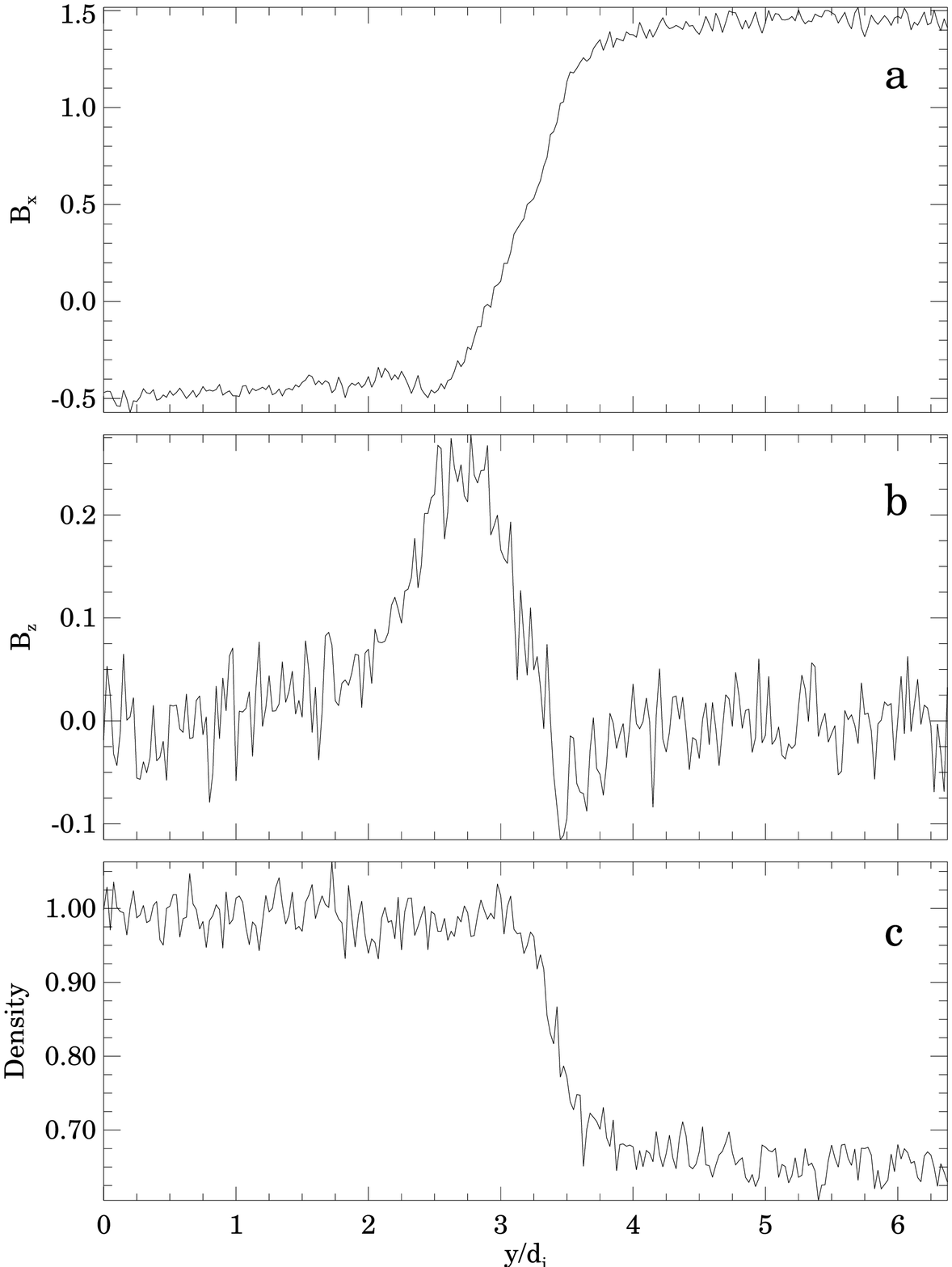}
\caption{\label{cutsfig} Cuts at $x = 20 d_i$ of the outflow region of
Figure \ref{psifig}. Each plot has been averaged over $0.25 d_i$ in
the horizontal direction to reduce the noise.  (a)-(c) The
reconnecting field, the out-of-plane field, and the density,
respectively.}
\end{figure}

Because of the simulation's relatively small size clearly separated
MHD and kinetic discontinuities do not exist, making comparisons with
previous work difficult. Nevertheless, in cuts of the reconnection
outflow (see Figure \ref{cutsfig}), the self-generated, out-of-plane
magnetic field is evident and the positive (negative) value of $B_z$
on the magnetosheath (magnetosphere) side of the current layer is
consistent with earlier predictions. The perturbation, however, is
much stronger on the magnetosheath side; the anti-symmetry of $B_z$
seen in systems with a symmetric pressure does not extend to
reconnection at the magnetopause.

The sharp density drop at the magnetosphere edge of the current layer
($y \approx 3.4 d_i$) is a consequence of the different inflow speeds
on the two sides of the layer and the plasma mixing across the island.
Since the magnetospheric field is much stronger, and therefore the
inflow velocity from this side of the current layer is weak, only a
small amount of magnetospheric plasma crosses the separatrices.  The
result is an outflow region almost completely comprising magnetosheath
plasma. This effect was evident in earlier hybrid simulations [{\it
Krauss-Varban et al.}  1999; {\it Nakamura and Scholer}, 2000] and has
also been seen in satellite crossings of the magnetopause [{\it
Eastman et al.}, 1996]\nocite{eastman96a}.

\subsection{Component Reconnection and Diamagnetic Propagation}\label{driftsec}

In the presence of a magnetic field any nonparallel pressure gradient
produces a diamagnetic drift,
\begin{equation}\label{vstar}
\mathbf{v}_{*j} = -c\frac{\bm{\nabla} p_j \bm{\times}
\mathbf{B}}{q_jnB^2}\text{,}
\end{equation}
where $p_j=nT_j$ is the thermal pressure and $q_j$ is the charge of
species $j$.  Due to the charge dependence, ions and electrons drift
in opposite directions.  To have diamagnetic flows parallel to the
current layer in our geometry the guide field $B_z$ and the gradient
of the pressure must be nonzero where $B_x = 0$.

Even though diamagnetic drifts do not correspond to actual particle
motions they nevertheless advect the magnetic field [{\it Coppi},
1965; {\it Scott and Hassam}, 1987]\nocite{coppi65a,scott87a}. In two
dimensions the magnetic field can be written as $\mathbf{B} =
\mathbf{\hat{z}} \bm{\times \nabla} \psi(x,y) +
B_z(x,y)\mathbf{\hat{z}}$ where $\psi$ is the magnetic flux
function. Taking the cross product of Faraday's law with
$\mathbf{\hat{z}}$ yields $\partial_t \bm{\nabla}\psi-c\bm{\nabla}
E_z=0$, or $E_z = \partial_t \psi/c$.  Next, dotting the electron
fluid equation with $\mathbf{\hat{z}}$ gives
\begin{equation}\label{ez}
E_z = -\frac{1}{c} \mathbf{\hat{z}} \cdot (\mathbf{v_e} \bm{\times}
\mathbf{B}) - \frac{m_e}{e}\frac{dv_{ez}}{dt}\text{.}
\end{equation}
The last term represents the inertial effects of electrons and breaks
the frozen-in condition.  Since it is usually small we ignore it and
substitute for $\mathbf{B}$ to get $E_z = -\mathbf{v_e} \bm{\cdot
\nabla} \psi/c$ and a convection equation for the flux [{\it Coppi},
1965],
\begin{equation}\label{convection}
\partial_t \psi + \mathbf{v_e} \bm{\cdot \nabla} \psi = 0 \text{.}
\end{equation}
Hence, the electron fluid velocity, which includes a diamagnetic
component given by (\ref{vstar}), advects magnetic structures.

\begin{figure}
\noindent\includegraphics[width=20pc]{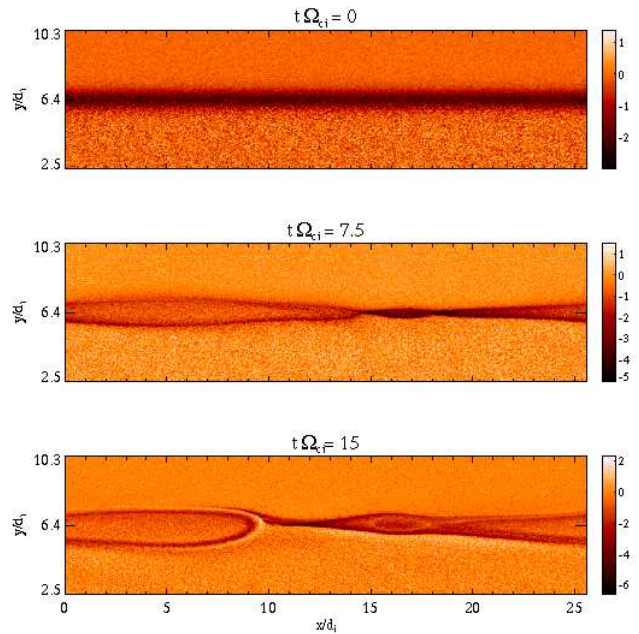}
\caption{\label{movefig} The out-of-plane current density at three
times for a magnetopause geometry similar to Figure \ref{psifig}, but
including an out-of-plane guide field and therefore an in-plane
diamagnetic drift.  The initial conditions are those described in
section \ref{inits} with an asymptotic guide field of $3.0$ in the
magnetosheath and $3.8$ in the magnetosphere. The density drops from
$1$ to $0.1$ across the current layer.}
\end{figure}

Strictly speaking, equations (\ref{vstar}) and (\ref{convection}) are
fluid results and need not describe the dynamics at the X-line of our
simulations where small-scale structures leave the fluid approximation
on unsure ground.  However, as can be seen in Figure \ref{movefig},
they are still good descriptions of the system.  The out-of-plane
current density essentially maps the magnetic field lines (see Figure
\ref{psifig}b) so the island (centered at $x/d_i \sim 5$) grows
robustly during the time shown. Simultaneously, the X-line propagates
to the left, the direction of the electron diamagnetic drift. The
locations of the X-line in the three panels of Figure \ref{movefig}
are $x/d_i = 19.2$, $14.8$ and $10$, respectively, corresponding to a
drift speed of $0.61 v_A$ and consistent with the value calculated
from the initial parameters of $v_{*e}=0.56$. As flux reconnects the
magnetic island widens, decreasing the pressure gradient across its
center, slowing its drift, and allowing the X-line to overtake
it. This stabilizes the island's growth because the plasma outflow
from the X-line towards the near end of the island must slow due to
the increase in magnetic pressure. The actual collision of the X-line
with the island, which occurs later in time, may be unphysical since
in any real system the initial separation of the X- and O-lines will
be much greater than the $12.8 d_i$ of the present model. Thus, the
effects of the collision will not be discussed further.

A surprise is that the diamagnetic propagation evident in Figure
\ref{movefig} has not been seen in previous hybrid simulations of the
magnetopause [{\it Krauss-Varban et al.}, 1999; {\it Nakamura and
Scholer}, 2000] even though these models should include the relevant
electron diamagnetic drifts.  In some cases the simulation parameters
may be such as to minimize diamagnetic effects ({\it e.g.} a lower
magnetospheric temperature).  Diamagnetic drifts may also be absent if
the simulations were performed with an initially uniform out-of-plane
field.  According to the force balance condition, the local pressure
gradient (and hence the drift) must then be zero at the reversal
surface, although whether it would remain zero as the system evolves
is unclear.  In section \ref{plane} we discuss further why a uniform
guide field may not be representative of reconnection at the
magnetopause.  A final possibility occurs for simulations using a
fixed, spatially-localized resistivity to break the frozen-in
condition at the X-line.  How such a resistivity model would affect an
X-line that is trying to propagate is unclear, but it is nevertheless
evident from our results that such models may be inappropriate for
simulating guide field reconnection at the magnetopause when
diamagnetic drifts are large.

\begin{figure}
\noindent\includegraphics[width=20pc]{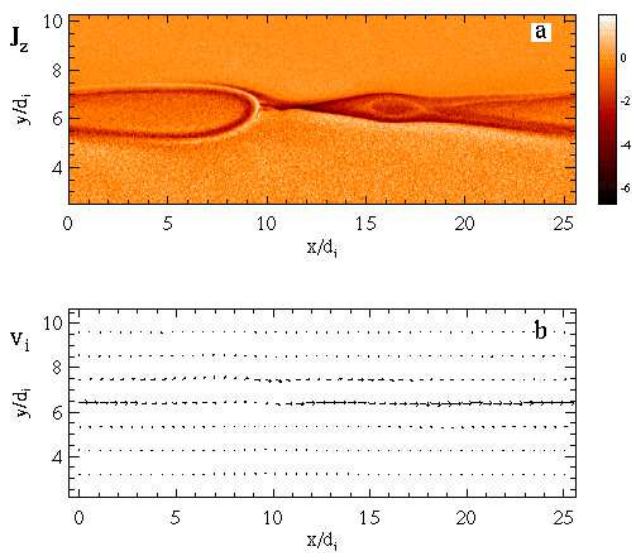}
\caption{\label{stabjzfig} Structure of (a) the out-of-plane current
density $J_z$ and (b) ion flow vectors at $t=16.0\Omega_{ci}^{-1}$ for
the simulation shown in Figure \ref{movefig}. The large flows at the
current layer are the ion diamagnetic drifts. The longest vector
corresponds to a velocity of $\sim 1.2 v_A$}
\end{figure}

Figure \ref{stabjzfig}a shows the current density for the simulation
of Figure \ref{movefig}.  Unlike the example shown in Figure
\ref{psifig}, the magnetic island does not bulge into the
magnetosphere side of the current layer.  This is because the
reconnecting magnetic field $B_x$ is anti-symmetric across the current
layer and thus, as discussed earlier, both sides contribute equal
amounts of magnetic flux to the reconnection.  Clearly evident in
Figure \ref{stabjzfig}a, as well as the last two panels of Figure
\ref{movefig}, is the left-right asymmetry of the opening angle of the
magnetic field lines near the X-line.  This is caused by the interplay
of the X-line's diamagnetic drift, controlled by just the electrons,
and the reconnection outflow, which must also involve the ions.

Consider the relative motion of the X-line and the ions.  Downstream
from the reconnection site the $\mathbf{J} \bm{\times} \mathbf{B}$
force accelerates frozen-in ions up to the Alfv\'en speed, $\pm
v_{Ax}$.  To this must be added the relative velocity between the
diamagnetic drift of the ions and the X-line, $v_*
=|v_{*i}|+|v_{*e}|$, with the result being a right-left asymmetry in
the outflow velocities, $\pm v_{Ax}+ v_*$.  Since, from continuity,
the opening angle of the magnetic field downstream from the X-line is
roughly $v_{in}/v_{out}$ --- and $v_{in}$ is roughly constant due to
the symmetric reconnecting field --- we expect the rightward opening
angle to be much smaller than its counterpart, consistent with Figure
\ref{stabjzfig}b.

By a similar argument, normal ion outflow from an X-line only develops
when $v_{Ax}>v_*$. For the simulations presented in Figures
\ref{movefig} and \ref{stabjzfig} the electron and ion diamagnetic
drifts at the reversal surface are $-0.56$ and $1.12$, respectively,
compared with $v_{Ax}=1.35$ (based on a density of $0.55$,
corresponding to a mixture of the magnetopause and magnetosheath
plasma). Thus, for the parameters of this simulation, the reconnection
should be marginally unable to reverse the ion flow to the left of the
X-line. The ion flow vectors are shown in Figure \ref{stabjzfig}b at
the same time as the current density shown in Figure
\ref{stabjzfig}a. The strong ion flow on the right side of the X-line
results from the combination of the ion diamagnetic velocity with the
direct acceleration by the reconnected magnetic field. The
reconnection generated forces almost reverse the diamagnetic flow just
to the left of the X-line.

We again emphasize that it is the relative drift of the ions and
electrons, {\it i.e.} $v_* = |v_{*i}|+|v_{*e}|$, that must be smaller
than the in-plane Alfv\'en speed for normal reconnection to
proceed. Reconnection was suppressed in a nearly indistinguishable
fashion in a series of simulations with the same total temperature
$T_e+T_i$ (and therefore identical $v_*$) but differing values of
$T_e$ and $T_i$ ($T_i = 1/2, T_e = 5/2$ and $T_i = 5/2, T_e =
1/2$). Since $T_i \gg T_e$ at the magnetopause, stabilization can
still occur even if the drift of the X-line, which is due only to
$v_{*e}$, is negligible.

\begin{figure}
\noindent\includegraphics[width=20pc]{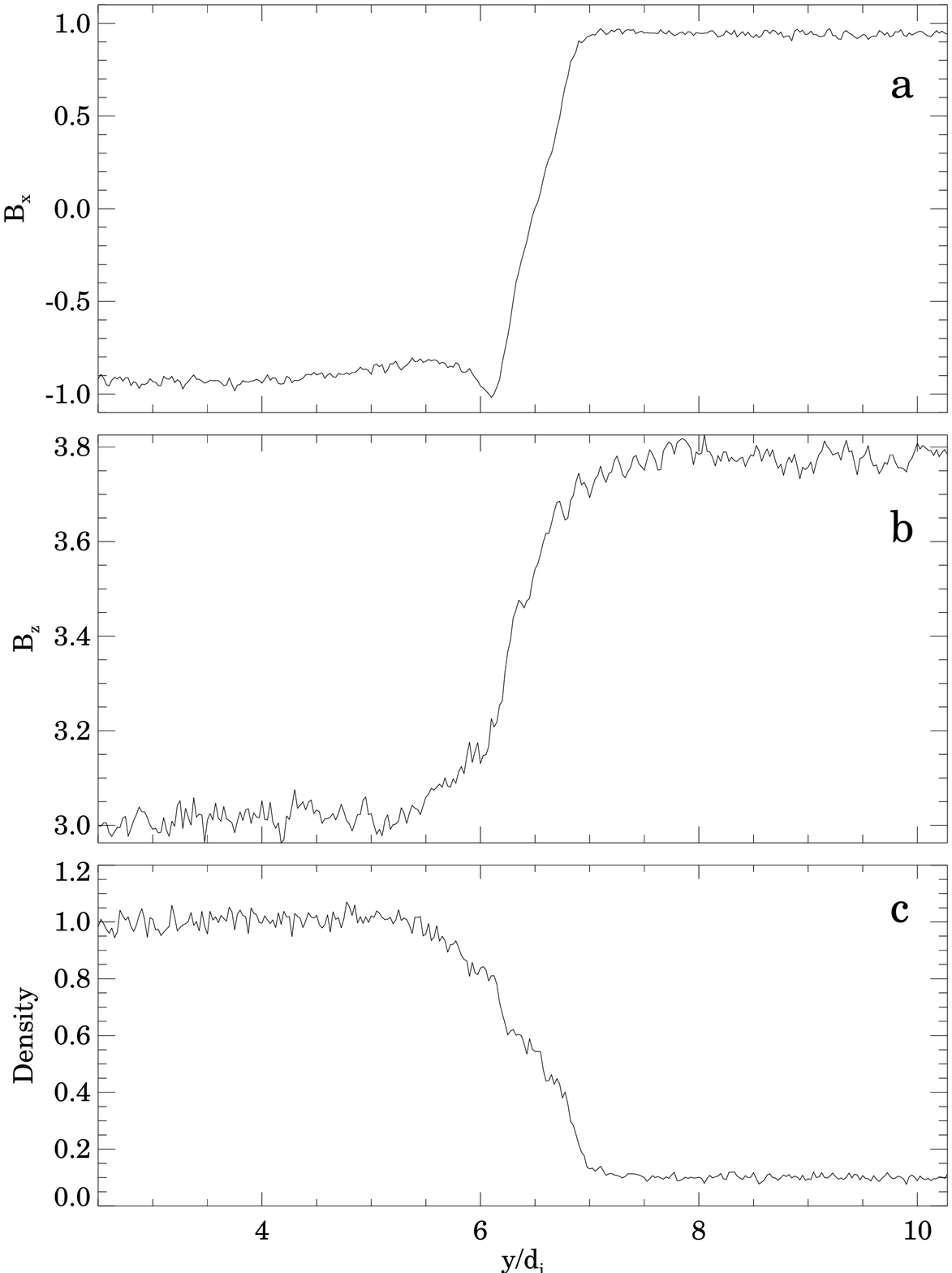}
\caption{\label{stabcuts} Cuts at $x = 13.8 d_i$ of the outflow region
of Figure \ref{stabjzfig}. Each plot has been averaged over $0.1 d_i$
in the horizontal direction to reduce the noise.  (a)-(c) The
reconnecting field, out-of-plane field and density, respectively.}
\end{figure}

The cuts in Figure \ref{stabcuts} show that for this simulation the
reconnection field $B_x$ is symmetric across the layer while the
pressure drop across the magnetopause is supported by the jump in the
guide field $B_z$. In contrast to the density profile in the case with
no guide field, the density drop across the layer is nearly centered
on the magnetic island. Large drops occur on either edge of the
current layer with a weak plateau in the middle of the island. The
inflow velocity into the island from either side of the current layer
is now roughly equal because of the near anti-symmetry of $B_x$. The
plasma in the core of the island is therefore a nearly equal mix of
magnetosphere and magnetosheath plasma and is not dominated by the
magnetosheath, in contrast to the case shown in section \ref{nodrift}.

\section{\label{betasec}Suppression of Component Reconnection}

\begin{figure}
\noindent\includegraphics[width=20pc]{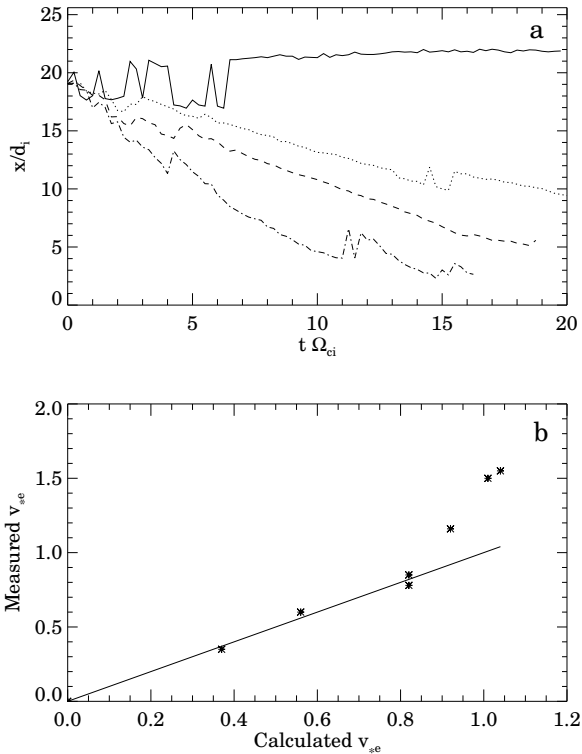}
\caption{\label{xlposfig} The top panel shows the position of the
X-line versus time for several runs differing only in the strength of
the guide field $B_z$.  The solid line is for the run of section
\ref{nodrift} and experiences no diamagnetic motion.  The dotted,
dashed, and dash-dotted lines correspond to $B_{z0} = 3, 1.5,$ and
$0.5$, respectively. The bottom panel is a comparison of the measured
X-line velocities and the electron diamagnetic velocity at the
reversal surface.  The straight line has a slope of one and is plotted
to guide the eye.  In both panels the X-point location corresponds to
a saddle point in $\psi$. }
\end{figure}

We have shown for a single case that, consistent with equation
(\ref{convection}), X-lines convect with the diamagnetic velocity of
the electrons in the reversal region and suggested a criterion for the
suppression of reconnection from diamagnetic drifts, $v_*>v_{Ax}$.  We
now more fully explore this stabilization criterion by varying the
strength of the out-of-plane field to vary the ratio of $v_*$ to
$v_{Ax}$. For our magnetopause model the diamagnetic velocity at the
surface where $B_x=0$ can be written as
\begin{equation}
v_{*j} =\frac{cT_j}{q_jL_{pj}B_z},
\end{equation}
where the pressure scale length is given by $L_p^{-1} =
|\partial_yp|/p$. Varying the amplitude of the out-of-plane field
$B_z$ while keeping all the other parameter fixed alters $v_{*j}$ but
keeps $v_{Ax}$ constant.  As can be seen in Figure \ref{xlposfig}a, a
large guide field implies a small $v_*$.  Despite the occasional
glitches (discontinuities in the motion arise when a new X-line
triggered by random perturbations begins to dominate), the drift
velocity is remarkably constant.

The bottom panel of Figure \ref{xlposfig} shows the initial electron
diamagnetic velocity at the reversal surface versus the drift speed of
the X-line measured in the simulation.  Although the agreement for
sub-Alfv\'enic velocities is quite good, as $v_{*e}$ approaches
$v_{Ax}$ another effect becomes important.  Rather than move the ions
at super-Alfv\'enic velocities the system develops an electric field
$E_y$ transverse to the current layer that, combined with the guide
field $B_z$, generates an $\mathbf{E} \bm{\times} \mathbf{B}$ drift
opposing the ion motion and adding to the electron velocity.  There is
no change in the net current but the electrons, and hence the X-line,
move faster than just the diamagnetic velocity would suggest.

\begin{figure}
\noindent\includegraphics[width=20pc]{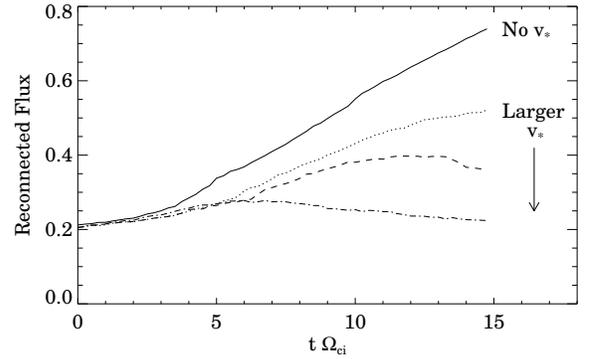}
\caption{\label{comp} Reconnected flux versus time for four
simulations.  The reconnection rate is the slope of each curve.  The
solid line is a reference run with no diamagnetic drift. The dotted,
dashed, and dash-dotted lines are from simulations with differing
values of $B_{z0}$ ($1.5, 1.0$ and $0$, respectively) and therefore
$v_*$.}
\end{figure}

Since diamagnetic flows of sufficient strength disrupt the large-scale
flows characteristic of reconnection, increases in $v_*$ should
correlate with decreases in the amount of reconnected flux. Figure
\ref{comp} depicts the extent to which a diamagnetic drift can alter
the reconnection rate of a system.  The reconnected flux is plotted
versus time for four different simulations: one with no diamagnetic
drift, as in Figure \ref{psifig}, and three with varying asymptotic
guide fields, $B_{z0} = 1.5, 1.0,$ and $0$.  For the latter cases the
associated drift speeds (diamagnetic plus $\mathbf{E} \bm{\times}
\mathbf{B}$) at the reversal surface are $v_*=0.8, 1.1$ and $1.5$,
respectively.  Reconnection is almost completely suppressed for the
largest drifts. The plateau in reconnected flux seen in each of the
bottom three curves is an artifact (discussed earlier) of our system's
periodic boundary conditions that allow the X-line to impinge on the
more slowly propagating magnetic islands.

For large guide fields the diamagnetic drift speed approaches zero and
the stabilization is minimized.  The converse limit is more subtle
since pressure balance imposes restrictions on the system's
configuration.  If the profiles of the density and the reconnecting
field are specified then there is a minimal possible value of the
guide field in the current layer; $B_z$ cannot be taken to be zero
and the drift cannot grow without bound.  Hence although (for
example) the GEM challenge simulations had zero guide field they did
not have a large diamagnetic drift.  Rather, $B_z\rightarrow 0$ and
$L_p \rightarrow \infty$, keeping $v_*=0$.

\begin{figure}
\noindent\includegraphics[width=20pc]{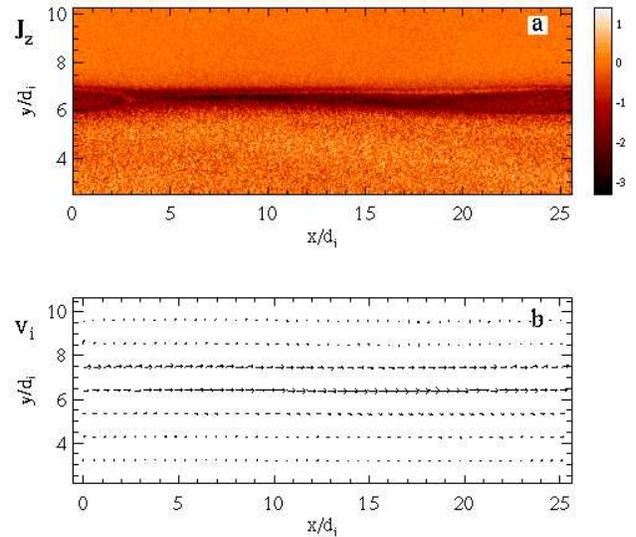}
\caption{\label{stabjzfig2} (a) The out-of-plane current density $J_z$
and (b) ion flow vectors at $t=13.0\Omega_{ci}^{-1}$ from a simulation
with an asymptotic $B_z=0.5$ and a strong diamagnetic drift. The large
flows to the right at the current layer are the result of the ion
diamagnetic drift. The longest vector corresponds to a velocity of
$\sim 0.95 v_A$}.
\end{figure}

To confirm the hypothesis that the reconnection process cannot reverse
the ion diamagnetic drift and drive outward flows from the X-line for
large values of the diamagnetic velocity, we show $J_z$ and the ion
velocities from a simulation with an asymptotic out-of-plane field
$B_z=0.5$ in Figure \ref{stabjzfig2}.  In this case the island of
reconnected flux remains small even at late time because of the weak
growth shown in Figure \ref{comp}.  The flows are actually smaller
than the peak of the expected diamagnetic velocity because of the
development of the electric field $E_y$ across the layer. A slight
reduction of the ion flow speed to the left of the X-line indicates
that the reconnected field lines exert a small leftward directed force
on the ions in this region, but the result is only a small
perturbation of the ion flow pattern. Thus, the X-line is unable to
couple effectively to the ions, strong reconnection does not develop,
and the island amplitude saturates.

\section{\label{plane}Dependence on the Reconnection Plane}

Although two-dimensional simulations are computationally cheaper than
their three-dimensional counterparts, this simplicity comes with
drawbacks.  Perhaps the most important when considering diamagnetic
drifts is that the orientation of the X-line with respect to the
magnetic field configuration is externally imposed, rather than being
free to develop.  Different simulation planes have different
reconnecting and out-of-plane fields and hence varying amounts of
diamagnetic stabilization.  In this section we explore this effect,
albeit only by considering rotations around the
$\mathbf{\hat{y}}$-axis to avoid introducing stabilizing $B_y$
components into the system.  We will assume that the favored plane
({\it i.e.}, the one the system would choose if allowed to evolve in
three dimensions) is the one where the reconnection rate is maximal,
but note that the results are not a replacement for more ambitious 3-D
simulations.  We further assume that the diamagnetic effects are
playing the dominant role in determining the plane where reconnection
is most robust.  Thus, we are considering systems in which $v_*\sim
v_{Ax}$.

Three-dimensional particle simulations have already addressed this
issue for the basic case of a system with no guide field and no
pressure drop across the current layer [{\it Hesse et al.}, 2001; {\it
Pritchett}, 2001].  Random perturbations were used as seeds to avoid
imposing a plane {\it a priori}.  Although X-lines were free to
develop with any orientation, they preferentially formed parallel to
the current layer, consistent with our assumption of treating only
rotations about $\mathbf{\hat{y}}$.  Moreover they remained
quasi-two-dimensional even at late times.  However, the simplicity of
$\mathbf{B}$ makes this case degenerate since the reversal surface
(the place where $B_x = 0$) must be at the center of the current layer
for any rotation about $\mathbf{\hat{y}}$.  This restriction does not
exist in more complicated configurations.

\begin{figure}
\noindent\includegraphics[width=20pc]{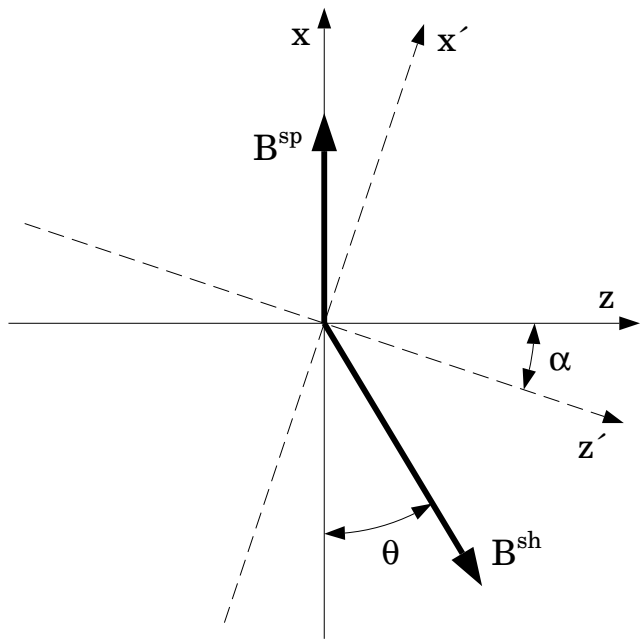}
\caption{\label{schematic} Schematic of the asymptotic magnetic fields
in the magnetosphere, $B^{sp}$, and magnetosheath, $B^{sh}$. The
reconnecting and out-of-plane components of the magnetic field depend
on the orientation of the reconnection plane: $x-z$ versus $x'-z'$.}
\end{figure}

The basic geometry of magnetic fields at the magnetopause is shown in
Figure \ref{schematic}. In order to agree with our simulation
coordinates, the magnetospheric field $\mathbf{B}^{sp}$ is taken to be
in the positive $x$ direction while the magnetosheath field
$\mathbf{B}^{sh}$ is inclined at an angle $\theta$ with respect to the
negative $x$ axis. Thus, the rotation angle across the magnetopause is
$\pi-\theta$.  The $\mathbf{\hat{y}}$ direction is perpendicular to
the surface of the magnetopause.  For a two-dimensional simulation in
the $x-y$ plane the reconnecting components of the magnetic field will
be parallel to $\mathbf{\hat{x}}$,
\begin{gather}
B^{sp}_x=B^{sp}\\ B^{sh}_x=-B^{sh}\cos\theta \text{,}
\end{gather}
and the guide field to $\mathbf{\hat{z}}$,
\begin{gather}
B^{sp}_z=0\\
B^{sh}_z=B^{sh}\sin\theta \text{.}
\end{gather}
The primed axes in Figure \ref{schematic} are rotated clockwise by an
angle $\alpha$ with respect to the unprimed system and represent
another possible reconnection plane.  In a simulation carried out in
the $x'-y'$ plane the reconnecting components of the magnetic field
will be
\begin{gather}
B^{sp}_{x'}=B^{sp}\cos\alpha \label{reconnection1}\\
B^{sh}_{x'}=-B^{sh}\cos(\theta+\alpha) \label{reconnection2} \text{,}
\end{gather}
while the guide field components are
\begin{gather}
B^{sp}_{z'}=-B^{sp}\sin\alpha\\ B^{sh}_{z'}=B^{sh}\sin(\theta+\alpha)
\text{.}
\end{gather}

\begin{figure}
\noindent\includegraphics[width=20pc]{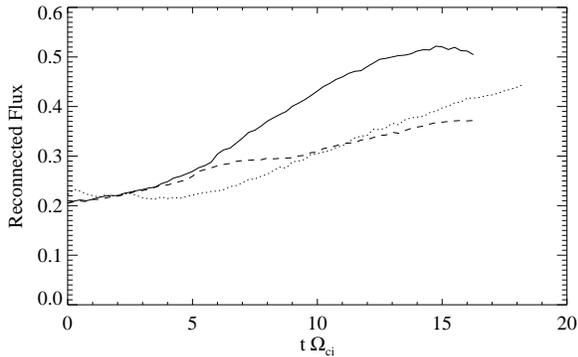}
\caption{\label{rotrate} Reconnected flux versus time for three
simulations.  The solid line is a simulation with an asymptotic guide
field of $1.5$ (the dotted line in Figure \ref{comp}).  The dashed and
dotted lines are simulations in planes rotated $\pm 15^{\circ}$ about
$\mathbf{\hat{y}}$.}
\end{figure}

As shown in section \ref{betasec}, stabilization via a diamagnetic
drift is inversely proportional to the value of the guide field $B_z$
in the center of the current layer. Thus, the expectation is that the
plane with the maximum value of the guide field at the surface where
reconnection takes place will dominate. Unfortunately, the asymptotic
fields determine neither the location of the reversal surface nor the
value of the guide field there.  Furthermore, the actual guide field
at the reversal surface may vary as reconnection proceeds.  To
simplify matters, we adopt the crude hypothesis that the guide field
at the reversal surface $B_g$ is the mean of the asymptotic guide
fields, which is $[B^{sh}\sin(\theta+\alpha)-B^{sp}\sin\alpha]/2$ in
the rotated system. With this {\it ansatz} the optimal plane for
reconnection is that where $B_g$ is maximized with respect to the
angle $\alpha$, from which we derive the condition
\begin{equation}
B^{sp}\cos\alpha=B^{sh}\cos(\theta+\alpha)\text{.}
\end{equation}
Interestingly, this corresponds to the plane where the reconnecting
field components are equal in magnitude but oppositely directed (see
equations (\ref{reconnection1})-(\ref{reconnection2})).  Figure
\ref{rotrate} shows the effect of rotating our simulation plane on the
reconnection rate and suggests that when diamagnetic drifts are
strong, reconnection in the anti-symmetric plane is favored.

\section{\label{conclusions}Summary and Discussion}

Diamagnetic drifts can be an important part of magnetopause
reconnection in certain parameter regimes.  In our simulations the
plasma $\beta$ (approximately $1$) and the magnetosheath-magnetosphere
density contrast (a factor of 10) are roughly consistent with
observational constraints.  However our initial conditions also
neglect both temperature gradients and scale lengths larger than an
ion inertial length.  For other magnetopause models that include these
features the diamagnetic drifts and associated stabilization may be
smaller.

When diamagnetic drifts are significant an X-line will advect with the
electron fluid velocity, causing the separatrices to have an asymmetry
in their opening angles with the wider facing the direction of motion.
For large enough flows, $v_*=|v_{*e}|+|v_{*i}|\sim v_A$, reconnection
can be completely suppressed since the large-scale outflows from the
X-line needed for fast reconnection cannot develop. The reconnected
field lines are not strong enough to reverse the ion diamagnetic flows
toward the X-line.

The diamagnetic effects presented here introduce a $\beta$ dependence
that was absent in the symmetric systems studied earlier. The
diamagnetic stabilization condition $v_*>v_A$ can be rewritten as a
condition on $\beta$:
\begin{equation}\label{beta}
\beta_x>\frac{B_z}{B_x}\frac{2L_p}{c/\omega_{pi}},
\end{equation}
where $B_x$ is the amplitude of the reconnecting field, $\beta_x=8\pi
nT/B_x^2$, and $B_z$ and $L_p$ are evaluated at the current layer. For
$B_z\sim B_x$ and for pressure scale lengths of the order of the ion
inertial length $c/\omega_{pi}$, the threshold for the suppression of
reconnection is $\beta\sim 1$.

A statistical study of accelerated flow events was done by
\nocite{scurry94a} {\it Scurry et al.} [1994] to investigate the
dependence of dayside reconnection on magnetosheath $\beta$ and the
shear angle of the magnetic field across the magnetopause. They found
that for low values of $\beta$ the accelerated flow events spread over
a broad range of clock angles while at high values of $\beta$ they
were strongly correlated with large clock angles. The implication is
that magnetic reconnection in the presence of a significant guide
field is suppressed at high $\beta$.  Thus, these observations are
consistent with the criterion given by equation (\ref{beta}) for the
diamagnetic suppression of reconnection.

When $B_z\gg B_x$ equation (\ref{beta}) indicates that it is difficult to
stabilize reconnection with diamagnetic effects, just as is
demonstrated in section \ref{betasec}.  We do not address what happens
when the guide field becomes very large, but {\it Rogers et al.}
[2001]\nocite{rogers01a} examined the effect of $\beta$ on
reconnection in a symmetric system (no pressure drop across the
current layer) using theoretical arguments as well as numerical
simulations.  They argued that fast reconnection as described in
section \ref{simoverview} can take place whenever dispersive waves
({\it e.g.}, whistlers or kinetic Alfv\'en waves) exist in a system at
small spatial scales. Only under extreme conditions ($B_z\sim
B_x\sqrt{m_i/m_e}$, $\beta_x=8\pi nT/B_x^2\ll 1$) is reconnection
inhibited by a guide field $B_z$.

A final curious feature of equation (\ref{beta}) occurs in the limit
where $B_z$ is small, as the expression would seem to imply that
reconnection is inhibited even for low values of $\beta_x$. This is
not the case because there is a linkage between the pressure scale
length $L_p$ and the value of $B_z$ at the reversal surface. Since the
reconnection field $B_x=0$ at this location, pressure balance requires
that $p/L_p=B_zB_z'$ with $B_z'=|\partial_yB_z|$. This condition
implies that $B_zL_p=p/B_z'$, and so that if the right-hand side is to
remain finite then $L_p\rightarrow \infty$ as $B_z\rightarrow 0$.  In
the absence of a guide field the pressure gradient must be zero at the
reversal surface.  For nonconstant $B_z$, the minimal initial value of
the reversal guide field is determined by the details of the initial
conditions.

Unlike systems with simple reversed fields, reconnection at the
magnetopause is not limited to a single plane, a fact that has led to
the prediction that the magnetopause field might become stochastic
[{\it Galeev et al.}, 1986; {\it Lee et al.}, 1993].  Diamagnetic
drifts offer a way to trigger reconnection at multiple surfaces since
they can directly affect which planes dominate reconnection.  A rough
argument suggests that the maximal guide field (and hence the minimal
drift and minimal stabilization) occurs at the surface where the
components of the field that reconnect are equal and opposite.  Hence
we expect this to be the dominant location of ``component
reconnection'' at the magnetopause.

The direction of the X-line advection can in principle be predicted
from the local pressure gradients and magnetic fields.  Under
appropriate conditions this effect should be detectable by spacecraft
at the magnetopause, although definitive results are certainly more
likely with multiple spacecraft missions like Cluster or the future
Magnetospheric Multiscale Mission.

\begin{acknowledgments}
This work was supported in part by the NASA Sun Earth Connection
Theory and Supporting Research and Technology programs and by the NSF.
\end{acknowledgments}
\newpage

\bibliographystyle{agu}
\bibliography{paper}

\end{article}

\end{document}